\documentclass[11pt,onecolumn]{elsart}    

\usepackage{graphicx}          

\usepackage{subfigure}
\usepackage{psfrag}
\usepackage{amsmath} 
\usepackage{amssymb}  
\usepackage{bm}
\usepackage{multirow}
\usepackage{color,soul}

\usepackage{empheq}
\usepackage{xcolor}

\setulcolor{red}
\setstcolor{green}
\sethlcolor{yellow}

\newcommand{\rtn}{\mathbb{R}}
\newcommand{\ntn}{\mathcal{N}}

\usepackage{colortbl,dcolumn}
\usepackage[switch,pagewise]{lineno}
\begin{document}

\begin{frontmatter}
\runtitle{Edge Agreement with Quantized Measurements}  

\title{Edge Agreement of Multi-agent System with Quantized Measurements via the Directed Edge Laplacian}

\thanks[footnoteinfo]{This paper was not presented at any meeting or journal.
Corresponding author Xiangke Wang. Tel. +86-0731-84576455.
}

\author[NUDT]{Zhiwen Zeng}\ead{zhiwenzeng.nudt@gmail.com},    
\author[NUDT]{Xiangke Wang}\ead{xkwang@nudt.edu.cn},               
\author[NUDT]{Zhiqiang Zheng}\ead{zqzheng@nudt.edu.cn}

\address[NUDT]{College of Mechatronics and Automation, National University of Defense Technology,
410073, China}  

\begin{keyword}                           
Edge agreement, edge Laplacian,  multi-agent system, quantized measurements.                 
\end{keyword}                             

\begin{abstract}
This work explores the edge agreement problem of second-order nonlinear multi-agent system under quantized measurements. Under the edge agreement framework, we introduce an important concept about the \emph{essential edge Laplacian} and also obtain a reduced model of the edge agreement dynamics based on the spanning tree subgraph. The quantized edge agreement problem of second-order nonlinear multi-agent system is studied, in which both uniform and logarithmic quantizers are considered. We do not only guarantee the stability of the proposed quantized control law, but also reveal the explicit mathematical connection of the quantized interval and the convergence properties for both uniform and logarithmic quantizers, which has not been addressed before. Particularly, for uniform quantizers, we provide the upper bound of the radius of the agreement neighborhood and indicate that the radius increases with the quantization interval. While for logarithmic quantizers, the agents converge exponentially to the desired agreement equilibrium. In addition, we figure out the relationship of the quantization interval and the convergence speed and also provide the estimates of the convergence rate. Finally, simulation results are given to verify the theoretical analysis.
\end{abstract}

\end{frontmatter}

\section{Introduction}
\vspace{-2pt}
Graph theory contributes significantly in the analysis and synthesis of multi-agent systems, since it provides natural abstractions for how information are shared among agents in a network \cite{olfati2004consensus,ren2005consensus,yang2011constructing,yang2014global}.
Pioneering researches on edge agreement \cite{zelazo2011edge,zelazo2013performance} not only provide totally new insights that how the spanning trees and cycles effect the performance of the agreement protocol, but also set up a novel systematic framework for analysing multi-agent systems from the edge perspective. In our previous work \cite{zeng2015convergence}, the concept of edge Laplacian was extended to more general directed graphs and the classical input-to-state nonlinear control methods together with the recently developed cyclic-small-gain theorem were successfully implemented to drive multi-agent system to reach robust consensus.


Early efforts on multi-agent systems mainly focuses on the study with high accurate data exchanging among agents. However, it is hard to be guaranteed in the real digital networks when considering that communication channel has a limited bandwidth, and energy used for transmission is generally restrained. Frankly speaking, constraints on communication have a considerable impact on the performance of multi-agent system. To cope with the limitations, the measurement data are always processed by quantizers. In practice, to realize the quantized communication scheme, a encoder-decoder pair is employed. Generally, the quantized data is always encoded by the sender side before transmitting and dynamically decoded at the receiver side.
Recently, the gossiping algorithms \cite{lavaei2012quantized}, the coding/decoding schemes \cite{li2011distributed} and nonsmooth analysis \cite{liu2013continuous} have been proposed to solve the coordination control problem of  first-order multi-agent systems with quantized information.
However, as known that second-order multi-agent systems have significantly different coordination behaviour even if agents are coupled through similar topology. To the best of authors' knowledge, there are still little works explore the quantization effects on second-order dynamics. Considering different quantizers, \cite{liu2012quantization} studies the synchronization behaviour of mobile agents with second-order dynamics under \hl{an undirected graph topology}. The authors also point out that the quantization effects may cause undesirable oscillating behaviour under directed topology. To more recent literature \cite{chen2013quantized}, the authors address the quantized consensus problem of second-order multi-agent systems via sampled data under directed topology. Considering the fact that quantization introduces strong nonlinear characteristics such as discontinuity and saturation to the system, the control law designed for the ideal case \hl{may lead} to instability. The research on second-order \hl{multi-agent systems} in the presence of quantized measurements under directed topology is still open.

While the analysis of the node agreement (consensus problem) has matured, work related to the edge agreement has not been deeply studied yet. In this paper, we are going to explore the quantization effects on the edge agreement problem of second-order nonlinear \hl{multi-agent systems}. The main contributions \hl{are twofold}. First, by introducing the essential edge Laplacian, we highlight the role of the spanning tree subgraph and then we can obtain a reduced model of second-order edge agreement dynamics across the spanning tree under the edge agreement framework. Second, we propose a general analysis of the convergence properties for second-order nonlinear \hl{multi-agent systems} under quantized measurements. Unlike previous works \cite{liu2012quantization,chen2013quantized}, for uniform quantizers, we provide the explicit upper bound of the radius of the agreement neighborhood and also indicate that the radius increases with the quantization interval. While for logarithmic quantizers, the agents converge exponentially to the desired agreement equilibrium. Moreover, we also provide the estimates of the convergence rate as well as pointing out that the coarser the quantizer is, the slower the convergence speed.

The rest of the paper is organized as follows: preliminaries are proposed in Section 2. The quantized edge agreement with second-order nonlinear dynamics under directed graph is studied in Section 3. The simulation results are provided in Section 4 while the last section draws the conclusions.

\section{Basic Notions and Preliminary Results}\label{sec:basis}
The null space of matrix $A$ is denoted by $\mathcal{N}(A)$. Denote by $I_n$ the identity matrix and by $\bm{0}_{n}$ the zero matrix in $\rtn ^{n\times n}$. Let $\bm{0}$ be the column vector with all zero entries. Let $\mathcal G = \left( {\mathcal V ,\mathcal E} \right)$ be a digraph of order $N$ specified by a node set $\mathcal V$ and an edge set $\mathcal E \subseteq  \mathcal V   \times \mathcal V$ with size $L$. The set of neighbors of node $i$ is denoted by $\ntn_i  = \left\{ {j: e_k = (j,i)  \in \mathcal E } \right\}$. The adjacency matrix of $\mathcal G$ is defined as ${A}_\mathcal{G} = \left[ {a_{ij} } \right] \in \mathbb{R}^{N \times N}$ with nonnegative adjacency elements $a_{ij} > 0 \Leftrightarrow \left( {j, i} \right) \in \varepsilon$. The degree matrix $\Delta_\mathcal{G} = \left[\Delta_{ij}\right]$ is a diagonal matrix with $\left[\Delta_{ii}\right] = \sum\nolimits_{j = 1}^N {a _{ij},i = 1,2, \cdots,N}$, and the graph Laplacian of $\mathcal{G}$ is defined by $L_\mathcal{G}\left( \mathcal G \right)=\Delta_\mathcal{G}-{A}_\mathcal{G}$. Denote by $\mathcal{W}(\mathcal{G})$ the $L \times L$ diagonal matrix of $w_k$, for $k=1,2\cdots,L$, where $w_k =  a_{ij}$ represents the weight of $e_k = (j,i) \in \mathcal{E}$.
The incidence matrix $E\left( \mathcal G \right) \in \rtn^{N \times L} $ for a directed graph is a $\left\{ {0, \pm 1} \right\}$-matrix with rows and columns indexed by nodes and edges of $\mathcal G$ respectively. For edge ${e_k}=(j,i) \in \mathcal{E}$, $\left[{E\left( \mathcal G \right)} \right]_{jk} = +1$, $\left[{E\left( \mathcal G \right)} \right]_{ik} = -1$ and $\left[{E\left(\mathcal G \right)} \right]_{lk} = 0$ if $l \ne i,j$. The in-incidence matrix ${E_{\odot} \left( \mathcal{G} \right)} \in \rtn^{N \times L}$ is a $\{ 0, - 1\}$ matrix and for ${e_k}=(j,i) \in \mathcal{E}$, $\left[{E_\odot \left( \mathcal G \right)} \right]_{lk} = -1$ for $l=i$, $\left[{E_\odot \left(\mathcal G \right)} \right]_{lk} = 0$ otherwise. The weighted in-incidence matrix $E_ \odot^w(\mathcal G)$ is defined as $E_ \odot^w(\mathcal G) = {E_{\odot} \left( \mathcal{G} \right)}\mathcal{W}(\mathcal{G})$.
As thus, the graph Laplacian of $\mathcal{G}$ has the following expression \cite{zeng2015convergence}:
$L_{\mathcal{G}}({\mathcal G}) = E_\odot^w({\mathcal G}) E({\mathcal G})^T.$
The weighted edge Laplacian of a directed graph $\mathcal{G}$ can be defined as \cite{zeng2015convergence}
\begin{linenomath}
\begin{align}\label{align:edgelap}
L_e({\mathcal G})  := E({\mathcal G})^T E_\odot^w({\mathcal G}).
\end{align}
\end{linenomath}
A spanning tree $\mathcal{G}_{_\mathcal{T}} = \left( {\mathcal V ,\mathcal E_1} \right)$ of a directed graph $\mathcal G = \left( {\mathcal V ,\mathcal E} \right)$ is a directed tree formed by graph edges that connect all the nodes of the graph; a co-spanning tree $\mathcal{G}_{_\mathcal{C}}= \left( {\mathcal V , \mathcal E - \mathcal E_1} \right) $ of $\mathcal{G}_{_\mathcal{T}}$ is the subgraph of $\mathcal G$ having all the vertices and exactly those edges of $\mathcal G$ that are not in $\mathcal{G}_{_\mathcal{T}}$. Graph $\mathcal{G}$ is called 
\emph{quasi-strongly connected} if and only if it has a directed spanning tree \cite{Thulasiraman:11b}. A quasi-strongly connected directed graph $\mathcal{G}$ can be rewritten as a union form: $\mathcal{G} = \mathcal{G}_{_\mathcal{T}}  \cup \mathcal{G}_{_\mathcal{C}}$. In addition, according to certain permutations, the incidence matrix $E(\mathcal G)$ can always be rewritten as $ E(\mathcal G) = \left[ {\begin{matrix}{E_{_\mathcal{T} }(\mathcal G)} & {E_{_\mathcal{C}}(\mathcal G)} \end{matrix}} \right]$ as well. Since the co-spanning tree edges can be constructed from the spanning tree edges via a linear transformation \cite{zelazo2011edge}, such that
\begin{linenomath}
\begin{align}\label{T}
E_{_{\mathcal T}}\left( {{{\mathcal G}}} \right)T({\mathcal G}) = E_{_{\mathcal C}}\left( {\mathcal G} \right)
\end{align}
\end{linenomath}
with $T({\mathcal G}) = {\left( {E_{_{\mathcal T}}{{\left( {{{\mathcal G}}} \right)}^T}E_{_{\mathcal T}}\left( {{{\mathcal G}}} \right)} \right)^{ - 1}}E_{_{\mathcal T}}{\left( {{{\mathcal G}}} \right)^T}E_{_{\mathcal C}}\left( {{{\mathcal G}}} \right)$ and $rank(E\left(\mathcal{G} \right)) = N-1$ from \cite{Thulasiraman:11b}. We define
\begin{linenomath}
\begin{align}\label{R}
R\left( {\mathcal G} \right) = \left[ {\begin{matrix}
   I & {T({\mathcal G})}  \cr
 \end{matrix}}\right]
\end{align}
\end{linenomath}
and then we have
\begin{linenomath}
\begin{align} \label{ER}
E\left( {\mathcal G} \right) = E_{_{\mathcal T}}\left( {{{\mathcal G}}} \right)R\left( {\mathcal G} \right).
\end{align}
\end{linenomath}
The column space of $E(\mathcal{G})^T$ is known as the \emph{cut space} of ${\mathcal G}$ and the null space of $E({\mathcal G})$ is called as the \emph{flow space} of $E({\mathcal G})$. Additionally, the rows of $R\left( {\mathcal G} \right)$ form a basis of the cut space of and the rows of $ \left[ {\begin{matrix}-{T({\mathcal G})}^T  & I \end{matrix}}\right]$ form a basis of the flow space, respectively \cite{Thulasiraman:11b}.
\begin{lem}[\cite{zeng2015convergence}]\label{theorem:Laplacianeigequal}
For a quasi-strongly connected graph $\mathcal{G}$, the graph Laplacian $L_{\mathcal{G}}({\mathcal G})$ and the edge Laplacian $L_e({\mathcal G})$ have the same $N-1$ nonzero eigenvalues, which are all in the open right-half plane.
\end{lem}

\begin{lem}[\cite{zeng2015convergence}]\label{thm:zeroeigen}
For a general quasi-strongly connected graph $\mathcal{G}=\mathcal{G}_{_\mathcal{T}} \cup \mathcal{G}_{_\mathcal{C}}$, $L_e({\mathcal G})$ contains $L-N+1$ zero eigenvalues. Moreover, if the edge set of $\mathcal{G}_{_\mathcal{C}}$ is not empty, then zero is a simple root of the minimal polynomial of $L_e({\mathcal G})$.
\end{lem}

\section{ Quantized Edge Agreement with Second-order Nonlinear Dynamics under Directed Graph}
In this section, the edge agreement of second-order nonlinear multi-agent systems under quantized measurements is studied.
To ease the notation, we simply use $E$, $E_ \odot^w$ and $L_e$ instead of  $E(\mathcal{G})$, $E_ \odot^w(\mathcal{G})$ and $L_e(\mathcal{G})$.

\subsection{Problem Formulation}
We consider a group of N networked agents and the dynamics of the $i$-th agent is represented by
\begin{linenomath}
\begin{align}
& \dot x_i(t)  = v_i(t) \label{dynamicsorder1} \\
& \dot v_i(t)  = f \left( {x_i(t) ,v_i(t),t } \right) +  u_i(t) \label{dynamicsorder2}
\end{align}
\end{linenomath}
where $x_i(t)  \in \rtn^n$ is the position, $v_i(t)  \in \rtn^n$ is the velocity and $u _i(t) \in \rtn^n$ is the control input. The nonlinear term  $f \left( {x_i(t) ,v_i(t),t } \right) :\rtn^n  \times \rtn^n  \to \rtn^n$ is unknown and \hl{satisfies} the following assumption:
\begin{assum}\label{assum}
For a nonlinear function $f$, there exists nonnegative constants $\xi_1$ and $\xi_2$ such that
\begin{linenomath}
\begin{align*}
\left| {f\left( {x,v,t} \right) - f\left( {y,z,t } \right)} \right| \le & \xi_1 \left| {x  - y } \right|+ \xi_2 \left| {v  - z } \right|,  \nonumber\\
& \forall x,v,y,z  \in {\rtn}^n ; \forall t \ge 0.
\end{align*}
\end{linenomath}
\end{assum}
The goal for designing distributed control law $u _i(t)$ is to synchronize velocities and positions of the $N$ networked agents.

The generally studied second-order consensus protocol proposed in \cite{yu2010second} is described as follows: $ u_i(t) =  \alpha \sum\limits_{j \in \mathcal{N}_i}^N {a_{ij}}\left( {{x_j}\left( t \right) - {x_i}\left( t \right)} \right) + \beta \sum\limits_{j \in \mathcal{N}_i}^N {{a_{ij}}\left( {{v_j}\left( t \right) - {v_i}\left( t \right)} \right)}$, for $i =1,2\cdots,N$, where $\alpha > 0$ and $\beta > 0$ are the coupling strengths. As in \cite{dimarogonas2010stability}, we assume that each agent $i$ has only quantized measurements of relative position $Q\left( {{x_i} - {x_j}} \right)$ and velocity information $Q\left( {{v_i} - {v_j}} \right)$, where $Q\left(.\right):\rtn^n \to \rtn^n$ denotes the \emph{quantization function}. Therefore, the protocol can be modified as
\begin{linenomath}
\begin{align}\label{quantizedpro}
u_i(t) = & \alpha \sum\limits_{j \in \mathcal{N}_i}^N {a_{ij}}Q\left( {{x_j}\left( t \right) - {x_i}\left( t \right)} \right)
 + \beta \sum\limits_{j \in \mathcal{N}_i}^N {{a_{ij}}Q\left( {{v_j}\left( t \right) - {v_i}\left( t \right)} \right)}
\end{align}
\end{linenomath}
for $i =1,2\cdots,N$. 
In this paper, two typical quantization operators are considered: uniform and logarithmic quantizer. For a given ${\delta _u} > 0$, a uniform quantizer ${q_u}:\rtn \to \rtn$ satisfies $\left| {{q_u}\left( a \right) - a} \right| \le {\delta _u},\forall a \in \rtn$; for a given ${\delta _l} > 0$, a logarithmic quantizer ${q_l}:\rtn \to \rtn$ satisfies $\left| {{q_l}\left( a \right) - a} \right| \le {\delta _l}\left| a \right|,\forall a \in \rtn $. The positive constants ${\delta _u}$ and ${\delta _l}$ are known as quantization interval. For a vector $\nu=[\nu_1,\nu_2,\cdots,\nu_n]^T\subset\rtn^n$, Let $Q_u\left(  \nu   \right) \buildrel \Delta \over = {\left[ {{q_u}\left(  \nu_1  \right),{q_u}\left(  \nu_2  \right), \cdots ,{q_u}\left(  \nu_n  \right)} \right]^T}$ and $Q_l\left(  \nu   \right) \buildrel \Delta \over = {\left[ {{q_l}\left( \nu_1  \right),{q_l}\left(  \nu_2  \right), \cdots ,{q_l}\left(  \nu_n  \right)} \right]^T}$. Then we obtain the following bounds: $\left| {{Q_u}\left( \nu  \right) - \nu } \right| \le \sqrt n {\delta _u}$, $\left| {{Q_l}\left( \nu  \right) - \nu } \right| \le {\delta _l}\left| \nu  \right|$.

Considering the dynamics of the networked agents described in \eqref{dynamicsorder1} and \eqref{dynamicsorder2}, by directly applying the quantized protocol \eqref{quantizedpro}, we obtain
\begin{linenomath}
\begin{align*}
\begin{cases}
{{\dot x}_i}\left( t \right) = {v_i}\left( t \right) \\
{{\dot v}_i}\left( t \right) =  f\left( {{x_i}\left( t \right),{v_i}\left( t \right),t} \right) +  \alpha \sum\limits_{j \in \mathcal{N}_i}^N {a_{ij}}Q\left( {{x_j}\left( t \right) - {x_i}\left( t \right)} \right) \nonumber \\
~~~~~~~ +  \beta \sum\limits_{j \in \mathcal{N}_i}^N {{a_{ij}}Q\left( {{v_j}\left( t \right) - {v_i}\left( t \right)} \right)}
\end{cases}
\end{align*}
\end{linenomath}
To ease the difficulty of the analysis, we technically chose $\alpha = \sigma^2$ and $\beta =\sigma^3$ ($\sigma > 0$) as in \cite{hu2012second}. The biggest advantage to using this trick is that we can easily construct a positive definite matrix which will be used in the proof of the main results. As thus, the system can be collected as
\begin{linenomath}
\begin{equation}\label{dyn1}
\begin{cases}
\dot x\left( t \right)  =  v\left( t \right)  \\
\dot v\left( t \right)  =  F\left( {x\left( t \right),v\left( t \right),t} \right) -  \sigma^2 ({E_ \odot^w } \otimes {I_n}) \hat Q\left( {({E^T}  \otimes {I_n})x\left( t \right)} \right)  \\
~~~~~~- \sigma^3 ({E_ \odot^w } \otimes {I_n})\hat Q\left( {({E^T} \otimes {I_n})v\left( t \right)} \right)
\end{cases}
\end{equation}
\end{linenomath}

with $x(t)$, $v(t)$ and $F( {x(t),v(t),t})$ denoting the column stack vector of ${x_i(t)}$, ${v_i(t)}$ and $f\left( {x_i(t),v_i(t),t} \right)$; and $\hat Q$ represents the vector form of the quantization function $Q$.

Define $x_e = (E^T \otimes {I_n}) x$ and $v_e = (E^T\otimes {I_n}) v$, which denote the difference of position and velocity of two neighbouring nodes respectively. We suppose ${e_{{x_e}}} = \hat Q\left( {{x_e}} \right) - {x_e}$ and ${e_{{v_e}}} = \hat Q\left( {{v_e}} \right) - {v_e}$ as in \cite{dimarogonas2010stability}. Then by left-multiplying $E^T\otimes {I_n}$ of both sides of \eqref{dyn1},
we have
\begin{linenomath}
\begin{equation}\label{edgesys}
\begin{cases}
{{\dot x}_e}\left( t \right) = {v_e}\left( t \right)  \\
{{\dot v}_e}\left( t \right) = (E^T\otimes {I_n}) F - \sigma^2 ({L_e} \otimes {I_n}){x_e} -\sigma^3 ({L_e} \otimes {I_n}){v_e} \\
~~~~~~~~  - \sigma^2 ({L_e} \otimes {I_n}){e_{{x_e}}} - \sigma^3 ({L_e} \otimes {I_n}){e_{{v_e}}}.
\end{cases}
\end{equation}
\end{linenomath}
The \emph{edge agreement dynamics} \eqref{edgesys} describes the evolution of the edge states $z = \left[ {\begin{matrix}
   x_e^T \quad v_e^T
  \end{matrix}} \right]^T$, which depends on its current state and its adjacent edges' states. In comparison to the node agreement (consensus), the edge agreement, rather than requiring the convergence to the agreement subspace, expects the edge dynamics \eqref{edgesys} to converge to the origin, i.e., $\mathop {\lim }\nolimits_{t \to \infty } \left| {{x_e}\left( t \right)} \right| = 0$ and $\mathop {\lim }\nolimits_{t \to \infty } \left| {{v_e}\left( t \right)} \right| = 0$.
\subsection{Main Results}
For the quasi-strongly connected graph $\mathcal{G}$, the incidence matrix can be written as $ E= \left[ {\begin{matrix}{E_{_\mathcal{T} }} & {E_{_\mathcal{C}}} \end{matrix}} \right]$.
Let $z_{_{\mathcal T}} = \left[ {\begin{matrix}
   x^T_{_{\mathcal T}} & v^T_{_{\mathcal T}} \cr
  \end{matrix}} \right]^T$ denotes the states across the spanning tree $\mathcal{G}_{_\mathcal{T}}$ with $x_{_{\mathcal T}} =  (E_{_\mathcal{T} }^T \otimes {I_n}) x$, $v_{_{\mathcal T}} =  (E_{_\mathcal{T} }^T \otimes {I_n}) v$ and $z_{_{\mathcal C}} = \left[ {\begin{matrix}
   x^T_{_{\mathcal C}} & v^T_{_{\mathcal C}} \cr
  \end{matrix}} \right]^T$ denotes the states across the cos-spanning tree $\mathcal{G}_{_\mathcal{C}}$ with $x_{_{\mathcal C}} = (E_{_\mathcal{C} }^T \otimes {I_n}) x$, $v_{_{\mathcal C}} =  (E_{_\mathcal{C} }^T \otimes {I_n}) v$, respectively. Notice that $E_{_{\mathcal T}}T(\mathcal G) = E_{_{\mathcal C}}$ as mentioned in \eqref{T}; therefore the co-spanning tree states can be reconstructed through matrix $T$, i.e., $ {x_{_{\mathcal C}}}\left( t \right) = ({T(\mathcal G)^T} \otimes {I_n}){x_{_{\mathcal T}}}\left( t \right)$ and $ {v_{_{\mathcal C}}}\left( t \right) = ({T(\mathcal G)^T} \otimes {I_n}){v_{_{\mathcal T}}}\left( t \right)$. 
  \hl{Moreover, based on the observation that $E = E_{_{\mathcal T}}R\left( {\mathcal G} \right)$ form} \eqref{ER}, \hl{then we can obtain $x_e = {{R^T(\mathcal G)}}\otimes {I_n} x_{_{\mathcal T}}$, $v_e = {{R^T(\mathcal G)}}\otimes {I_n} v_{_{\mathcal T}}$ and}
\begin{linenomath}
\begin{align}\label{xext}
z = \left[ {\begin{matrix}
   {{R^T(\mathcal G)}}\otimes {I_n} & \bar{\bm{0}}  \cr
   \bar{\bm{0}} & {{R^T(\mathcal G)}}\otimes {I_n}  \cr
 \end{matrix}}  \right] {z_{_{\mathcal T}}}
\end{align}
\end{linenomath}
\hl{with zero matrix $\bar{\bm{0}}$ of compatible dimension.}

To simplify the subsequent analysis, the essential edge Laplacian will be employed, which helps us to obtain a reduced model of the closed-loop multi-agent system based on the spanning tree subgraph $\mathcal{G}_{_\mathcal{T}}$.

Before moving on, we introduce the following transformation matrix:
\begin{linenomath}
\begin{align*}
{S_e}\left( {\mathcal G} \right) = \left[ \begin{matrix}
   {R{{\left( {\mathcal G} \right)}^T}} & {{\theta_e}\left( {\mathcal G} \right)}  \cr
 \end{matrix}  \right]~~
{S_e}{\left( {\mathcal G} \right)^{ - 1}} = \left[ {\begin{matrix}
   {{{\left( {R\left( {\mathcal G} \right)R{{\left( {\mathcal G} \right)}^T}} \right)}^{ - 1}}R\left( {\mathcal G} \right)}  \cr
   {{\theta_e}\left( {\mathcal G} \right)^T}  \cr
 \end{matrix}}  \right]
\end{align*}
\end{linenomath}
where $R\left( {\mathcal G} \right)$ is defined via \eqref{R} and  ${{\theta_e}\left( {\mathcal G} \right)}$ denotes the orthonormal basis of the flow space, i.e., $E{{\theta_e}\left( {\mathcal G} \right)} = 0$. Since $rank(E) = N-1$, one can obtain that $dim({\theta_e}\left( {\mathcal G} \right))= \mathcal{N}(E)$ and ${\theta_e}\left( {\mathcal G} \right)^T{\theta_e}\left( {\mathcal G} \right)=I_{L-N+1}$. Applying the above similar transformation lead to
\begin{linenomath}
\begin{align}\label{simtrans}
{S_e}{\left( {\mathcal G} \right)^{ - 1}}{L_e}{S_e}\left( {\mathcal G} \right) = \left[ {\begin{matrix}
   {{\hat L}_e} & E_{_{\mathcal T}}^T{E_ \odot^w }{{\theta_e}\left( {\mathcal G} \right)}  \cr
   \bar{\bm{0}} & \bar{\bm{0}} \cr
 \end{matrix}}  \right]
 \end{align}
\end{linenomath}
where ${{\hat L}_e} = E_{_{\mathcal T}}^T{E_ \odot^w }{R(\mathcal G)^T}$ is \hl{referred} to as the essential edge Laplacian. For the essential edge Laplacian, we have the following lemma.
\begin{lem}
The essential edge Laplacian ${{\hat L}_e}$ contains exactly $N-1$ nonzero eigenvalues of $L_e$.
\end{lem}
\begin{pf}
Based on the similar transformation ${S_e}\left( {\mathcal G} \right)$ and ${S_e}{\left( {\mathcal G} \right)^{ - 1}}$, the eigenvalues of the block matrix \eqref{simtrans} are the solution of
\begin{linenomath}
\begin{align*}
\lambda^{(L-N+1)} \det \left( {\lambda I - {{\hat L}_e}} \right) = 0
\end{align*}
\end{linenomath}
which shows that ${{\hat L}_e}$ contains exactly all the nonzero eigenvalues of $L_e$ from Lemma \ref{theorem:Laplacianeigequal} and \ref{thm:zeroeigen}.
\end{pf}
Meanwhile, we can construct the following Lyapunov equation {as
\begin{linenomath}
\begin{align}\label{edgelyap}
 \colorbox{yellow}{$\ H{{\hat L}_e} + \hat L_e^TH = -I_{N - 1}$}
\end{align}
\end{linenomath}
where $H$ is positive definite.

Next, we will provide a reduced multi-agent system model in terms of the corresponding dynamics across $\mathcal{G}_{_\mathcal{T}}$. For edge dynamics \eqref{edgesys}, we make use of the following transformation
\begin{linenomath}
\begin{align*}
(S_e^{ - 1}\otimes {I_n}){x_e} =  \left( {\begin{matrix}
    {{x_{_{\mathcal T}}}}  \cr
    \bm{0}  \cr
 \end{matrix} } \right)~~~~
 (S_e^{ - 1}\otimes {I_n}){v_e} =  \left( {\begin{matrix}
    {{v_{_{\mathcal T}}}}  \cr
    \bm{0}  \cr
 \end{matrix} } \right).
\end{align*}
\end{linenomath}
\hl{Since $E = E_{_{\mathcal T}}R\left( {\mathcal G} \right)$ and $L_e= E^T E_\odot^w$, we have} $S_e^{ - 1}E^{ T} = \left( {\begin{matrix}
    E_{_{\mathcal T}}^T \cr
    \bm{0}  \cr
 \end{matrix} } \right)$ \hl{and} $S_e^{ - 1}L_e = \left( {\begin{matrix}
    E_{_\mathcal{T}}^T E_\odot^w \cr
    \bm{0}  \cr
 \end{matrix} } \right)$. \hl{Then we define} ${\omega} = \left[ {\begin{matrix}
   {e_{x_e}}^T \quad {e_{v_e}}^T
  \end{matrix}} \right]^T$  \hl{and let} ${\hat L}_{_{\mathcal O}} = E_{_\mathcal{T}}^T E_\odot^w$. \hl{By using the similar transformation} \eqref{simtrans}, \hl{system} \eqref{edgesys} \hl{finally can be recast into a compact matrix form as follows:}
\begin{linenomath}
\begin{align}\label{STsubsys}
{{\dot z}_{_{\mathcal T}}} = {{\mathcal F}_{_{\mathcal T}}} +  ({{\mathcal L}_{_{\mathcal T}}} \otimes {I_n}){z_{_{\mathcal T}}} + ({ {{\mathcal L}_{_{\mathcal T 1}}}} \otimes {I_n}){\omega}
\end{align}
\end{linenomath}
\hl{with} ${{\mathcal L}_{_{\mathcal T}}} = \left[ {\begin{matrix}
   {{\bm{0}_{N - 1}}} & {{I_{N - 1}}}  \cr
   { - \sigma^2 {{\hat L}_e}} & { - \sigma^3 {{\hat L}_e}}  \cr
  \end{matrix} }  \right]$, ${{\mathcal L}_{_{{\mathcal T}1}}} = \left[ {\begin{matrix}
   {{\bm{0}_{N-1\times L}}} & {{\bm{0}_{N-1 \times L}}}  \cr
   { - \sigma^2 {{\hat L}_{_{\mathcal O}} }} & { - \sigma^3 {{\hat L}_{_{\mathcal O}} }}  \cr
  \end{matrix}} \right]$ \hl{and} $ {\mathcal F}_{_{\mathcal T}} = \left[{\begin{matrix}
   {{\bm{0}}}  \cr
    \cellcolor{yellow!20}{(E_{_{\mathcal T}}^T\otimes {I_n}){F}}  \cr
  \end{matrix} }\right]$.

\begin{rem}
The decomposition of the spanning tree and co-spanning tree subgraph has been wildly applied to solve many magnetostatic problems, such as tree-cotree gauging \cite{manges1995generalized}, finite element analysis \cite{lee2008application}. As is well known, the spanning tree \hl{plays} a vital role in the stability analysis of networked multi-agent system. Under the edge agreement framework, we reveal the connection of the algebraic properties and the graph structure and highlight the role of the spanning tree subgraph by introducing the essential edge Laplacian.
\end{rem}

To further look at the relation between the quantization interval and the edge agreement, we propose the following theorem.
\begin{thm}\label{the:main}
Considering the quasi-strongly connected directed graph $\mathcal G$ associated with the edge Laplacian $L_e$ , suppose $\mathcal{Q} =  - \left( {{\mathcal P}{{\mathcal L}_{_{\mathcal T}}} + {\mathcal L}_{_{\mathcal T}}^T{\mathcal P}} \right)$ with ${\mathcal P} = \left[ {\begin{matrix}
   {\sigma H} & {{H}}  \cr
   {{H}} & {\sigma H}  \cr
   \end{matrix}} \right]$, where $H$ is obtained by \eqref{edgelyap}. If $\sigma > \sqrt {{{{\lambda _{\max }(H)}} \over 2} + 1} $ and $\lambda_{\min } \left(\mathcal Q \right) -$ $2max\left( {{\xi _1},{\xi _2}} \right)\left\| {{\mathcal P}} \right\| > 0$.
Then, under the quantized protocol \eqref{quantizedpro}, system \eqref{STsubsys} has the following convergence properties:

(1): With uniform quantizers, the agents converge to a ball of radius
\begin{linenomath}
\begin{align}\label{uccond}
\left| {{z_{_{_{\mathcal T}}}}} \right| \le {{2\sqrt {2nL} {\delta _u}\left\| {{\mathcal P}{{\mathcal L}_{_{{\mathcal T}1}}} } \right\|} \over {\lambda_{\min } \left(\mathcal Q \right) - 2max\left( {{\xi _1},{\xi _2}} \right)\left\| {{\mathcal P}} \right\|}}
\end{align}
\end{linenomath}
which is centred at the agreement equilibrium;

(2): With logarithmic quantizers, the agents converge exponentially to the desired agreement equilibrium, provided that ${\delta _l}$ satisfies
\begin{linenomath}
\begin{align}\label{lccond}
{\delta _l} < {{\lambda_{\min } \left( \mathcal Q \right) - 2 max\left( {{\xi _1},{\xi _2}} \right)\left\| {{\mathcal P} } \right\|} \over {2\left\| {{\mathcal P}{{\mathcal L}_{_{{\mathcal T}1}}} } \right\|\left\| {{R^T}} \right\|}}.
\end{align}
\end{linenomath}
The estimated trajectories of the edge Laplacian dynamics \eqref{STsubsys} is as
\begin{linenomath}
\begin{align*}
\left|z_{_{\mathcal T}}(t)\right| \le {{\lambda_{max}(\mathcal{P})} \over {\lambda_{min}(\mathcal{P})}}e^{-{{\pi}\over{\lambda_{\max}(\mathcal P)}} t}\left|{{z_{_{\mathcal T}}}(0)}\right| ~\text{for}~t \ge 0
\end{align*}
\end{linenomath}
with
\begin{linenomath}
\begin{align*}
\pi =   \lambda_{\min } \left( \mathcal Q \right) - 2 max\left( {{\xi _1},{\xi _2}} \right)\left\| {{\mathcal P} } \right\| -
 2{\delta _l}\left\| {{\mathcal P}{{\mathcal L}_{_{{\mathcal T}1}}} } \right\|\left\| {{R^T}} \right\|.
\end{align*}
\end{linenomath}
\end{thm}

\begin{pf}
For the edge Laplacian dynamics \eqref{STsubsys}, we choose the following Lyapunov function candidate:
\begin{linenomath}
\begin{align}\label{ly}
V\left( {{z_{_{\mathcal T}}}} \right) = z_{_{\mathcal T}}^T({\mathcal P} \otimes {I_n}){z_{_{\mathcal T}}}
\end{align}
\end{linenomath}
in which
$
{\mathcal P} = \left[ {\begin{matrix}
   {\sigma H} & {{H}}  \cr
   {{H}} & {\sigma H}  \cr
   \end{matrix}} \right]
$
, where $H$ can be obtained from \eqref{edgelyap} and $\mathcal P$ is positive definite while choosing $\sigma>1$.

By taking the derivative of \eqref{ly} along the trajectories of \eqref{STsubsys}, we have
\begin{linenomath}
\begin{align*}
\dot V\left( {{z_{_{\mathcal T}}}} \right) = & z_{_{\mathcal T}}^T\left( {{\mathcal P}{{\mathcal L}_{_{\mathcal T}}}\otimes {I_n} + {\mathcal L}_{_{\mathcal T}}^T{\mathcal P}}\otimes {I_n} \right){z_{_{\mathcal T}}} + 2z_{_{\mathcal T}}^T({\mathcal P}\otimes {I_n}){{\mathcal F}_{_{\mathcal T}}} \\
   &+ 2z_{_{\mathcal T}}^T({\mathcal P}{{\mathcal L}_{_{{\mathcal T}1}}}\otimes {I_n}){\omega} \\
=  & -z_{_{\mathcal T}}^T(\mathcal{Q}\otimes {I_n}){z_{_{\mathcal T}}} + 2z_{_{\mathcal T}}^T({\mathcal P}\otimes {I_n}){{\mathcal F}_{_{\mathcal T}}}  + 2z_{_{\mathcal T}}^T({\mathcal P}{{\mathcal L}_{_{{\mathcal T}1}}}\otimes {I_n}){\omega}
\end{align*}
\end{linenomath}
in which
\begin{linenomath}
\begin{align*}
\mathcal{Q} =  - \left( {{\mathcal P}{{\mathcal L}_{_{\mathcal T}}} + {\mathcal L}_{_{\mathcal T}}^T{\mathcal P}} \right) = \left[ \begin{matrix}
   {\sigma^2{I_{N - 1}}} & {{\sigma^3}{I_{N - 1}} - \sigma H}  \cr
   {{\sigma^3}{I_{N - 1}} - \sigma H} & {{\sigma^4}{I_{N - 1}}- 2H}  \cr
  \end{matrix} \right].
\end{align*}
\end{linenomath}
Let $\mathcal{Q} = \left[ {\begin{matrix}
   {{\mathcal{Q}_1}} & {{\mathcal{Q}_2}}  \cr
   {\mathcal{Q}_2^T} & {{\mathcal{Q}_3}}  \cr
 \end{matrix}} \right]$ with $\mathcal{Q}_1 = {\sigma^2{I_{N - 1}}}$, $\mathcal{Q}_2 = {\sigma^3}{I_{N - 1}} - \sigma H$ and $\mathcal{Q}_3 = {{\sigma^4}{I_{N - 1}}- 2H}$. According to Schur complements theorem \cite{yu2010second}, by selecting
\begin{linenomath}
 $$\sigma > \sqrt {{{{\lambda _{\max }(H)}} \over 2} + 1} $$
\end{linenomath}
then we have $\mathcal{Q}_1 > 0$ and $\mathcal{Q}_3 - \mathcal{Q}_2^T {{\mathcal{Q}_1}}^{-1}{{\mathcal{Q}_2}} = H\left( {2\left( {{\sigma ^2} - 1} \right){I_{N - 1}} - H} \right) >0$, so that $\mathcal{Q}$ is positive definite.

In the meanwhile, we notice that
\begin{linenomath}
\begin{align}\label{normf}
\left| {{{\mathcal F}_{_{\mathcal T}}}} \right|  = \left| {(E_{_{\mathcal T}}^T \otimes {I_n})\mathcal{F}} \right| \le
 max\left( {{\xi _1},{\xi _2}} \right)\left| {{z_{_{\mathcal T}}}} \right|.
\end{align}
\end{linenomath}

For uniform quantizers, we can calculate the upper bound of the quantization error as
\begin{linenomath}
\begin{align}\label{normque}
\left| {\omega} \right| & \le \sqrt{2nL}\delta _u.
\end{align}
\end{linenomath}
By combining \eqref{normf} and \eqref{normque}, one can obtain
\begin{linenomath}
\begin{align*}
\dot V\left( {{z_{_{\mathcal T}}}}  \right)
& \le  - \lambda_{\min } \left( \mathcal Q \right){\left| {{z_{_{_{\mathcal T}}}}} \right|^2} + 2 max\left( {{\xi _1},{\xi _2}} \right)\left\| {{\mathcal P} } \right\|{\left| {{z_{_{\mathcal T}}}} \right|^2} \\
& + 2\sqrt {2nL} {\delta _u}\left\| {{\mathcal P}{{\mathcal L}_{_{{\mathcal T}1}}} } \right\|\left| {{z_{_{\mathcal T}}}} \right|  \\
& = \left| {{z_{_{_{\mathcal T}}}}} \right|\Bigl( (- \lambda_{\min } \left(\mathcal Q \right) + 2 max\left( {{\xi _1},{\xi _2}} \right)\left\| {{\mathcal P} } \right\|)\left| {{z_{_{_{\mathcal T}}}}} \right| \\
& + 2\sqrt {2nL}{\delta _u}\left\| {{\mathcal P}{{\mathcal L}_{_{{\mathcal T}1}}} } \right\| \Bigr).
\end{align*}
\end{linenomath}
Clearly, the edge agreement can be reached and the radius of the agreement neighbourhood is as \eqref{uccond}.

For logarithmic quantizers, according to $\left| {{Q_l}\left( a\right) - a} \right| \le {\delta _l} \left|a\right|$ and equation \eqref{xext} , we have
\begin{linenomath}
\begin{align}\label{normqle}
\left| \omega \right| & \le {\delta _l}\left|z\right| \le {\delta _l}\left\| {{R^T}} \right\|\left| {{z_{_{\mathcal T}}}} \right|.
\end{align}
\end{linenomath}
Combining \eqref{normf} and \eqref{normqle}, we have
\begin{linenomath}
\begin{align*}
\dot V\left( {{z_{_{\mathcal T}}}}  \right)
& \le  - \lambda_{\min } \left( \mathcal Q \right){\left| {{z_{_{_{\mathcal T}}}}} \right|^2} + 2 max\left( {{\xi _1},{\xi _2}} \right)\left\| {{\mathcal P} } \right\|{\left| {{z_{_{\mathcal T}}}} \right|^2} \\
& + 2{\delta _l}\left\| {{\mathcal P}{{\mathcal L}_{_{{\mathcal T}1}}} } \right\|\left\| {{R^T}} \right\|{\left| {{z_{_{\mathcal T}}}} \right|^2}  \\
& = -\pi{\left| {{z_{_{_{\mathcal T}}}}} \right|^2}.
\end{align*}
\end{linenomath}
Obviously, while the condition \eqref{lccond} is satisfied, the edge Laplacian dynamics \eqref{STsubsys} converges exponentially to the desired agreement equilibrium.

Moreover, one can obtain that
\begin{linenomath}
\begin{align*}
\dot V\left( {{z_{_{\mathcal T}}}}(t)  \right)  \le -\pi{\left| {{z_{_{_{\mathcal T}}}}} \right|^2} \le -{{\pi}\over{\lambda_{\max}(\mathcal P)}} V\left( {{z_{_{\mathcal T}}}(t)}  \right).
\end{align*}
\end{linenomath}
By applying the Comparison Lemma \cite{khalil2002nonlinear}, we have
\begin{linenomath}
\begin{align*}
V\left( z_{_{\mathcal T}}(t)  \right)
& \le e^{-{{\pi}\over{\lambda_{\max}(\mathcal P)}} t}V\left( {{z_{_{\mathcal T}}}(0)}  \right).
\end{align*}
\end{linenomath}
Then we can provide the following estimates of the convergence rate for the closed-loop multi-agent system
\begin{linenomath}
\begin{align}\label{esm}
\left|z_{_{\mathcal T}}(t)\right| \le {{\lambda_{max}(\mathcal{P})} \over {\lambda_{min}(\mathcal{P})}}e^{-{{\pi}\over{\lambda_{\max}(\mathcal P)}} t}\left|{{z_{_{\mathcal T}}}(0)}\right| ~\text{for}~t \ge 0.
\end{align}
\end{linenomath}
\end{pf}
\begin{rem}
From equation \eqref{uccond}, one can see that the radius of the convergence neighbourhood trends to zero as the quantization interval $\delta_u$ \hl{decreases}. Additionally, for logarithmic quantizers, the corresponding convergence rate of the quantized system depend on $\delta_l$, \hl{i.e.,} the coarser the logarithmic quantizer is, more time it takes to converge. As the convergence time can be used to measure the performance of the quantized control law, we provide the estimation of the upper bound of the convergence time based on \eqref{esm} as follows:
\begin{linenomath}
\begin{align*}
 T =  - {{{\lambda _{\max }}\left( \mathcal P \right)} \over \pi }\ln {{{\lambda _{\min }}\left( \mathcal P \right)r} \over {{\lambda _{\max }}\left( \mathcal P \right)\left|{{z_{_{\mathcal T}}}(0)}\right| }}
\end{align*}
\end{linenomath}
where $r>0$ denotes the expected radius of the agreement error.
\end{rem}
\section{Simulation}
Consider multi-agent system consisting of a group of $5$ agents associated with a quasi-strongly connected graph as shown in Fig \ref{span}, where ${e_1},{e_2},{e_3},{e_4} \subset {{\mathcal G}_{_{\mathcal T}}}$ and ${e_5} \subset {{\mathcal G}_{_{\mathcal C}}}$.
\begin{figure}[hbtp]
\centering
{\includegraphics[height=3cm]{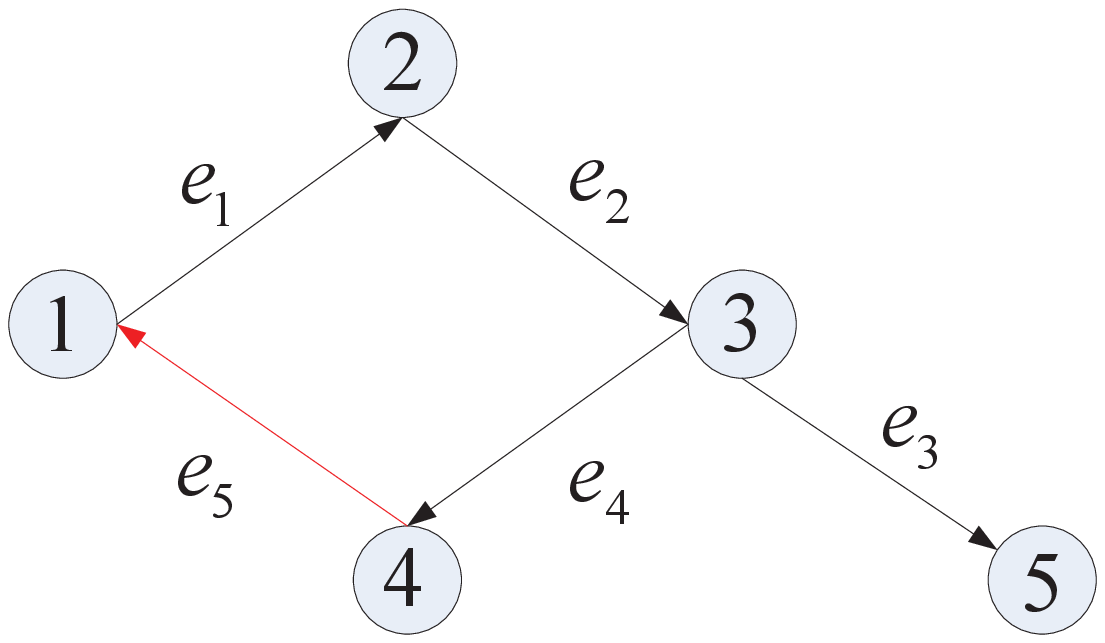}}
\caption{A quasi-strongly connected graph of $5$ agents.}
\label{span}
\end{figure}

The dynamics of the $i$-th agent is described as \eqref{dynamicsorder1} and \eqref{dynamicsorder2} with $x_i\left(t\right), v_i\left(t\right)$ $u_i\left(t\right) \in \rtn^3$. Let $x(m,:)$ and $v(m,:)$ denote the column vector of the $m$-th (m =1,2,3) variable of $x\left(t\right), v\left(t\right)$ respectively. The inherent nonlinear dynamics $f \left( {x_i(t) ,v_i(t),t } \right) :\rtn \times \rtn^3 \to \rtn^3$ is described by Chua's circuit
\begin{linenomath}
\begin{align*}
f \left( {x_i(t) ,v_i(t),t } \right) =
(\zeta \left( { - v_{i1}(t)  + v_{i2}(t)  - l\left( {v_{i1}(t) } \right)} \right), \nonumber\\
\tau (v_{i1}(t)  - v_{i2}(t)  + v_{i3}(t)) , - \chi v_{i2}(t) )^T
\end{align*}
\end{linenomath}
where $l\left( {v_{i1}(t) } \right) = bv_{i1}(t)  + 0.5\left( {a - b} \right)\left( {\left| {v_{i1}(t)  + 1} \right| - \left| {v_{i1}(t)  - 1} \right|} \right)$. The system is chaotic when $\zeta  = 0.01$, $\tau = 0.001$, $\chi  = 0.018$, $a =  - 4/3$ and $b =  - 3/4$. In view of Assumption \ref{assum}, simple calculation leads to $\xi_1 = 0$ and $\xi_2 = 4.3871\times 10^{-3}$ \cite{yu2010second}.

Suppose that the weighted diagonal matrix is $\mathcal{W}=diag\{0.12,0.24,0.44,0.43,0.09\}$. By choosing $\sigma = 1.64$, we have
\begin{linenomath}
$${\hat L}_e = {\begin{pmatrix}\begin{smallmatrix}
    0.21  &  0.09   &      0.00  &  0.09   \cr
   -0.12  &  0.24   &      0.00  &  0.00  \cr
    0.00  & -0.24  &  0.44   &      0.00   \cr
    0.00  & -0.24  &       0.00   & 0.43   \cr
 \end{smallmatrix}\end{pmatrix} },~
 {\hat L}_{_{\mathcal O}}=  {\begin{pmatrix}\begin{smallmatrix}
    0.12 &  0.00  &  0.00 &  0.00 &  -0.09 \cr
     -0.12 &   0.24 &  0.00  &  0.00  &  0.00 \cr
    0.00 &  -0.24  &  0.44 &  0.00  & -0.00 \cr
     0.00  & -0.24 &  0.00 &   0.43  &  0.00 \cr
 \end{smallmatrix}\end{pmatrix} } .
$$
\end{linenomath}
\subsection{Uniform Quantizer}
First, we consider the quantized protocol \eqref{quantizedpro} with the following uniform quantizer as the one used in \cite{liu2012quantization},
\begin{linenomath}
\begin{align}\label{qus}
{q_u}\left( x \right) = {\delta _u}\left( {\left\lfloor {{x \over {{\delta _u}}}} \right\rfloor  + {1 \over 2}} \right).
\end{align}
\end{linenomath}
The simulation results with $\delta_u = 1$ are shown in Fig. \ref{fig:qu}, from which we can see that
$x_e(t)$ and $v_e(t)$ indeed converge to a small neighbourhood near the equilibrium points. To show the effect of $\delta_u$ on the agreement error $\left|e_{ss}\right|$, we further take $\delta_u =0.01,0.1,2~\text{and}~3$ to run the simulation. The results in Tab.\ref{tb1} shows that error trends to zero when $\delta_u \to 0$,  \hl{as shown in} Theorem \ref{the:main}.

\begin{figure*}[hbtp]
\begin{center}
\mbox{
{\includegraphics[width=72mm]{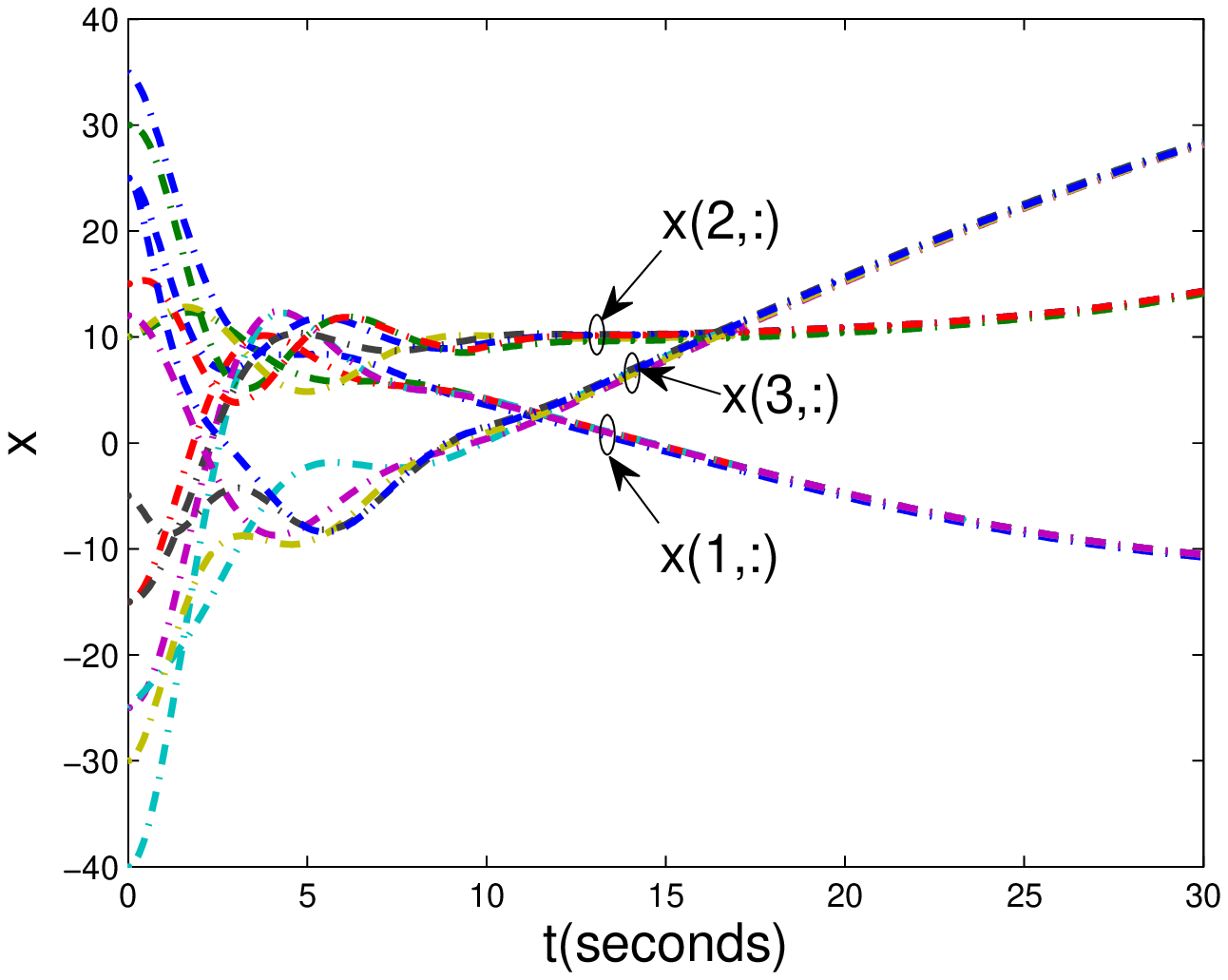}\label{fig:qux}}
{\includegraphics[width=72mm]{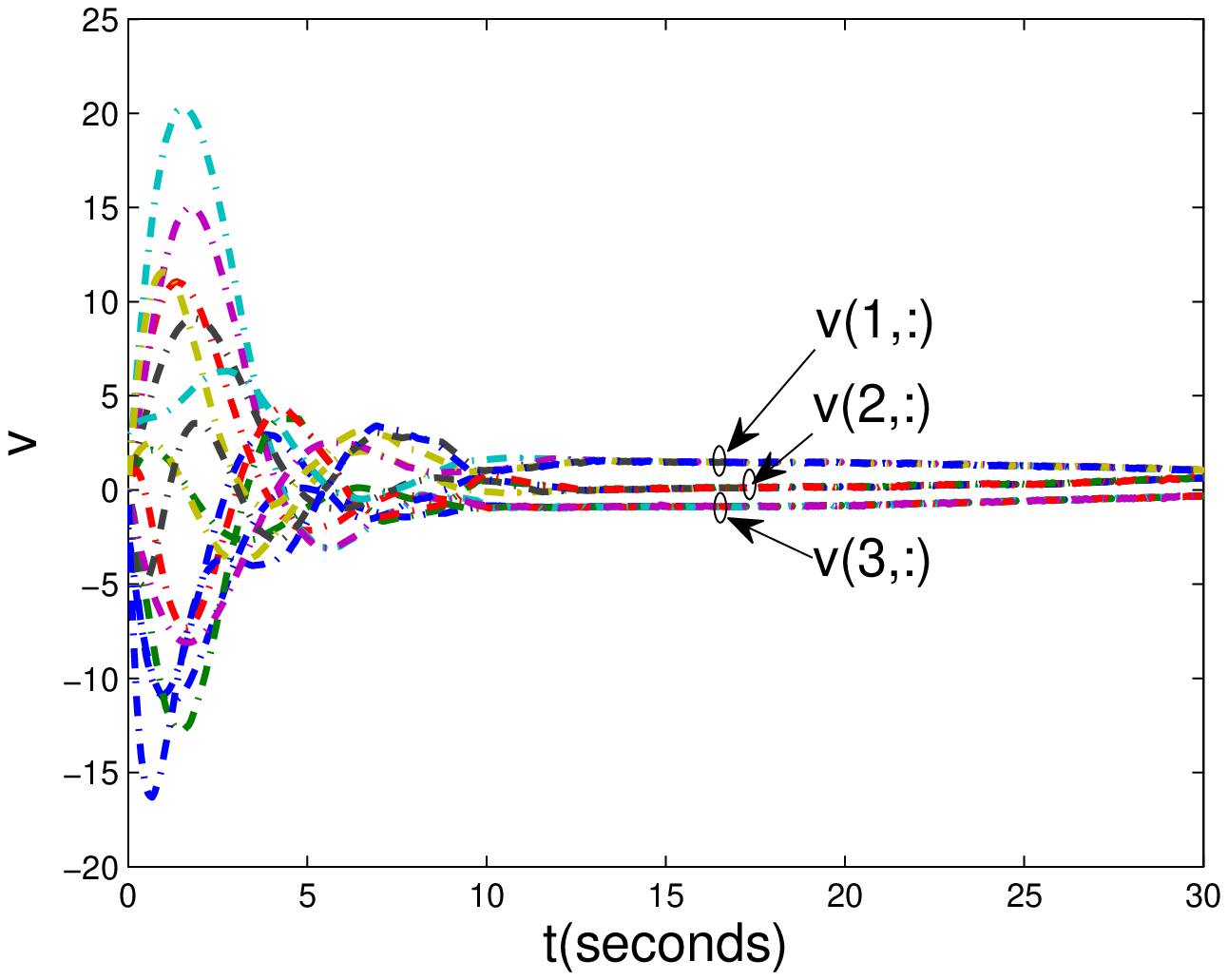}\label{fig:quv}}
}
\mbox{
{\includegraphics[width=72mm]{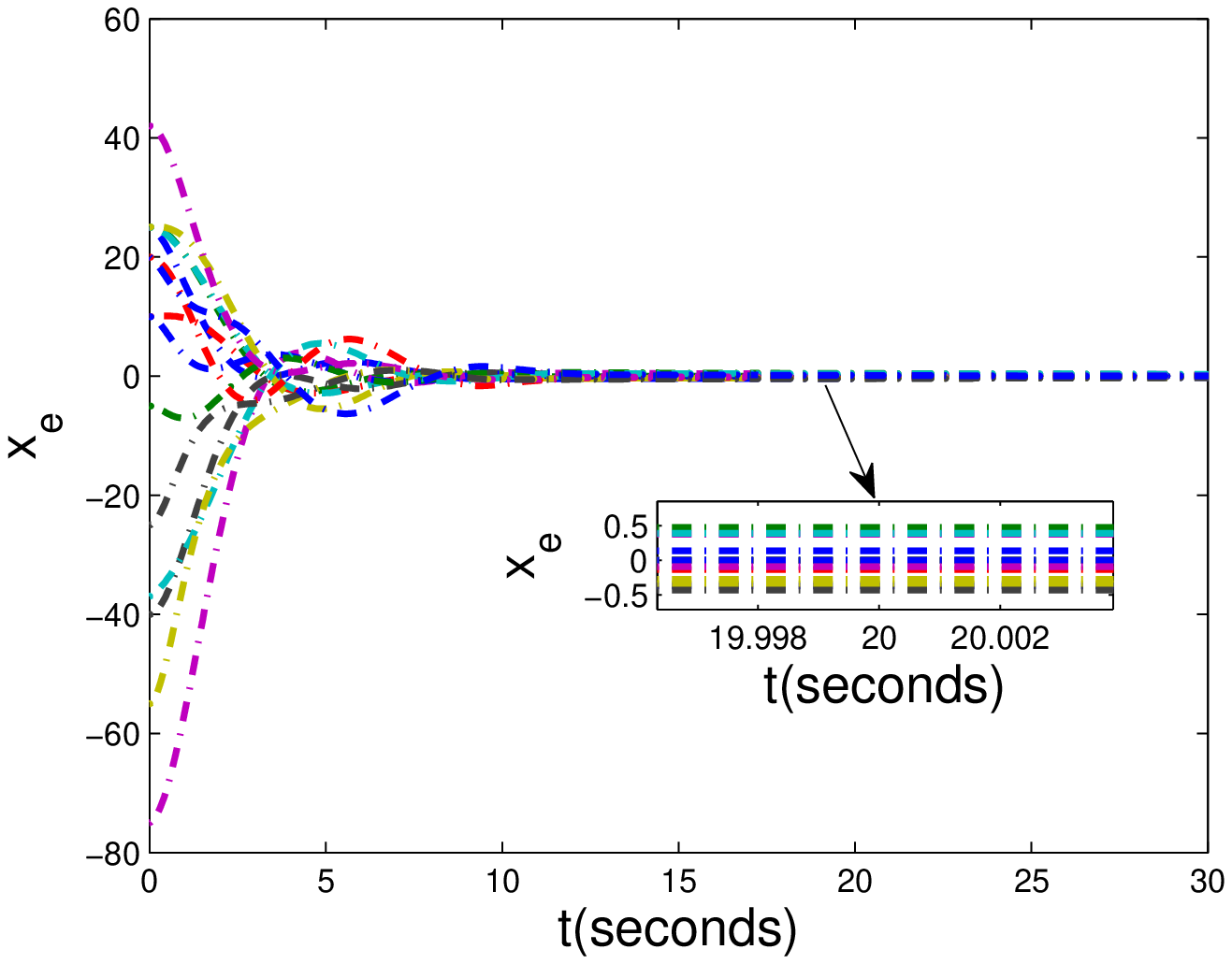}\label{fig:quxe}}
{\includegraphics[width=72mm]{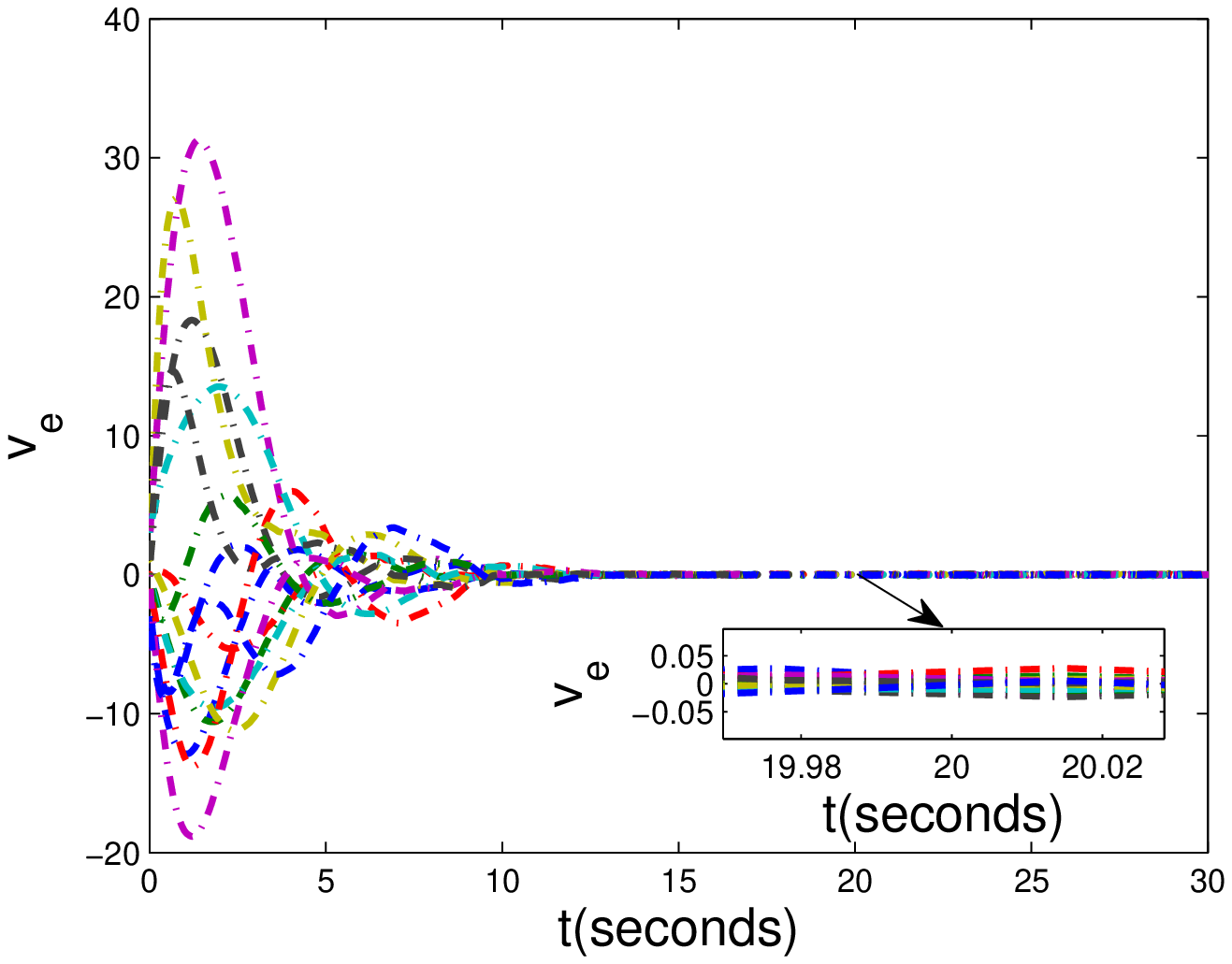}\label{fig:quve}}
}
\caption{Edge agreement under uniform quantizer with $\delta_u = 1$.}
\label{fig:qu}
\end{center}
\end{figure*}

\begin{table}[!hbp]
\begin{center}
\caption{The effect of $\delta_u$ on the agreement error.}
\begin{tabular}{|c|c|c|c|c|c|}
\hline
$\delta_u$  & 0.01 & 0.1 & 1 & 2 & 3 \\
\hline
error & 0.005 & 0.05 & 0.62 & 1.98 & 4.45 \\
\hline
\end{tabular}\label{tb1}
\end{center}
\end{table}
\subsection{Logarithmic Quantizer}
The logarithmic quantizer we apply to \eqref{quantizedpro} \hl{is an odd map ${q_l}:\rtn \to \rtn$} \cite{liu2012quantization},
\begin{linenomath}
\begin{align}
{q_l} = \left\{ {\begin{matrix}
   {{e^{{q_u}\left( {\ln x} \right)}}}~~~~~~~~\text{when $x>0$}  \cr
   0  ~~~~~~~~~~~~~~~~~\text{when $x=0$}\cr
   { - {e^{{q_u}\left( {\ln \left( { - x} \right)} \right)}}}~~\text{when $x<0$}  \cr
 \end{matrix}} \right.
\end{align}
\end{linenomath}
where ${q_u}$ is defined as \eqref{qus} and the parameter $\delta_l = 1-e^{-\delta_u}$. To satisfy the stability constraints \eqref{lccond}, we require $\delta_l < 0.0301$. The simulation results with $\delta_u = 0.01$ and $\delta_l = 1-e^{-0.01} =0.01$ are shown in Fig.\ref{fig:ql}, from which we can see that the agreement is indeed achieved as well as $x_i(t)$ and $v_i(t)$ converge to the desired agreement values. The estimation of the convergence rate is given as
$
\left|\psi\right|={{\lambda_{max}(\mathcal{P})} \over {\lambda_{min}(\mathcal{P})}}e^{-{{\pi}\over{\lambda_{\max}(\mathcal P)}} t}\left|{{z_{_{\mathcal T}}}(0)}\right| ~\text{for}~t \ge 0
$
with $\pi=0.5387$. From Fig.\ref{qlest}, one can see that $\left|z_{_\mathcal{T}}\right|$ exponentially converge to the origin. 
\begin{figure*}[hbtp]
\begin{center}
\mbox{
{\includegraphics[width=72mm]{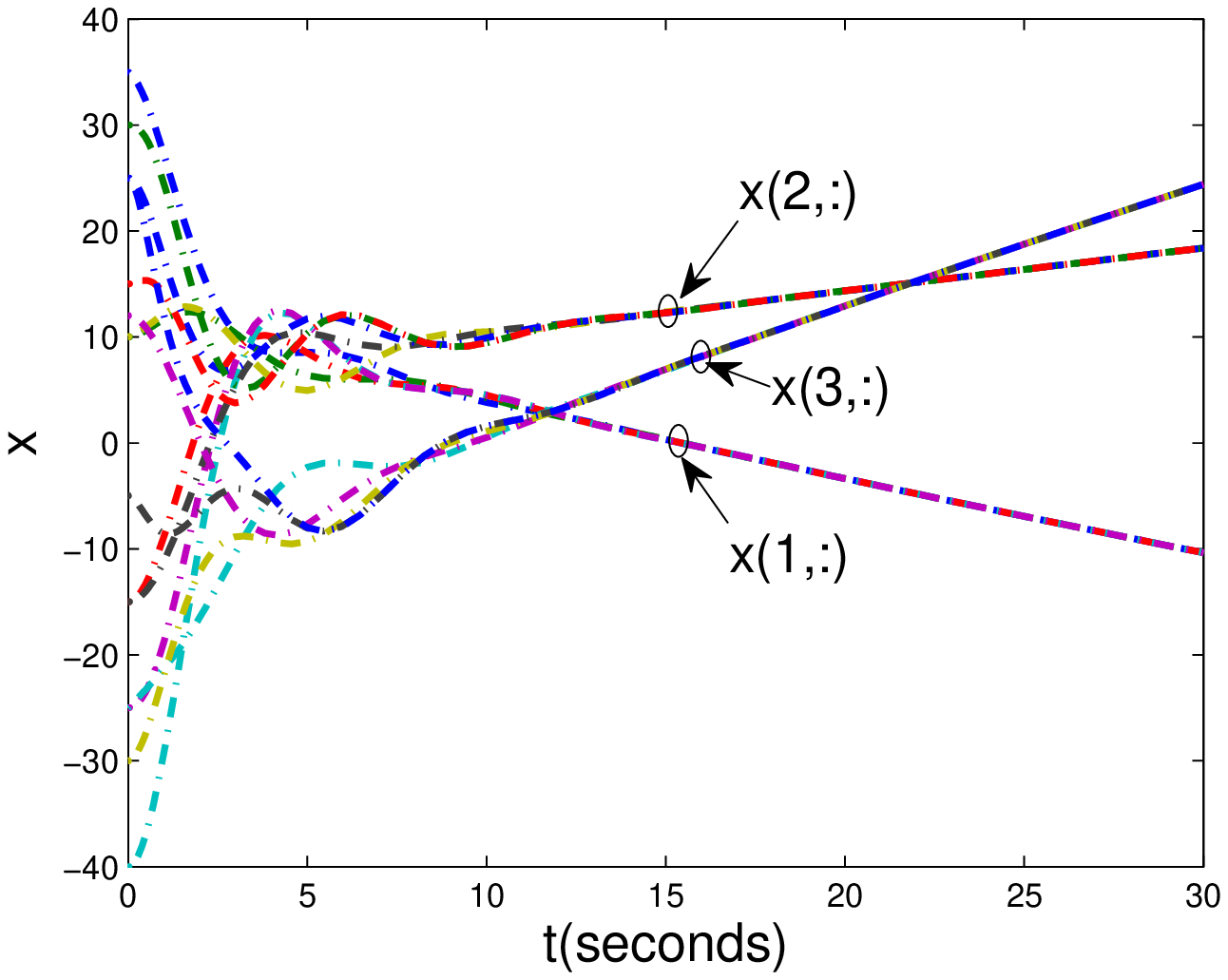}\label{fig:qlx}}
{\includegraphics[width=72mm]{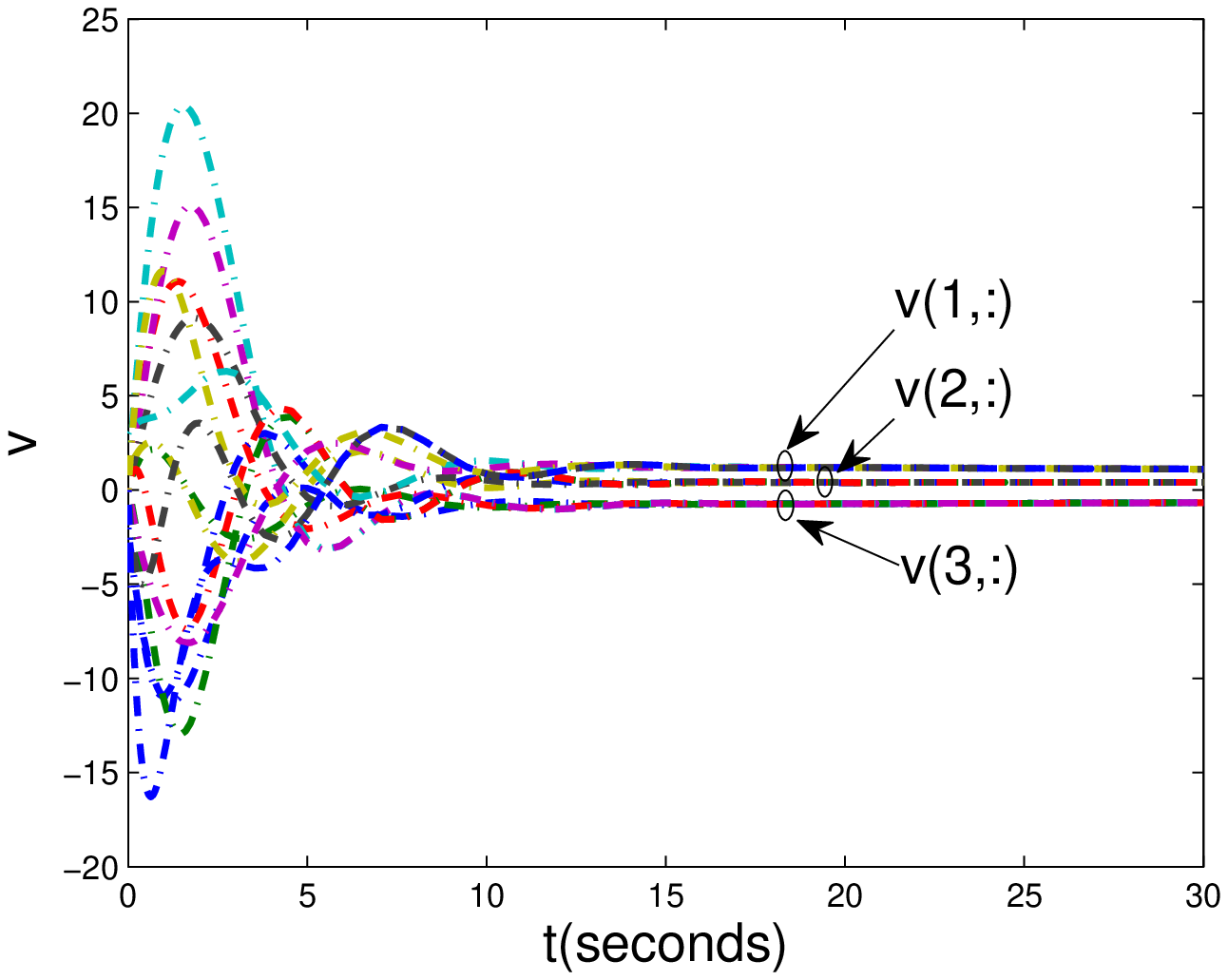}\label{fig:qlv}}
}
\mbox{
{\includegraphics[width=72mm]{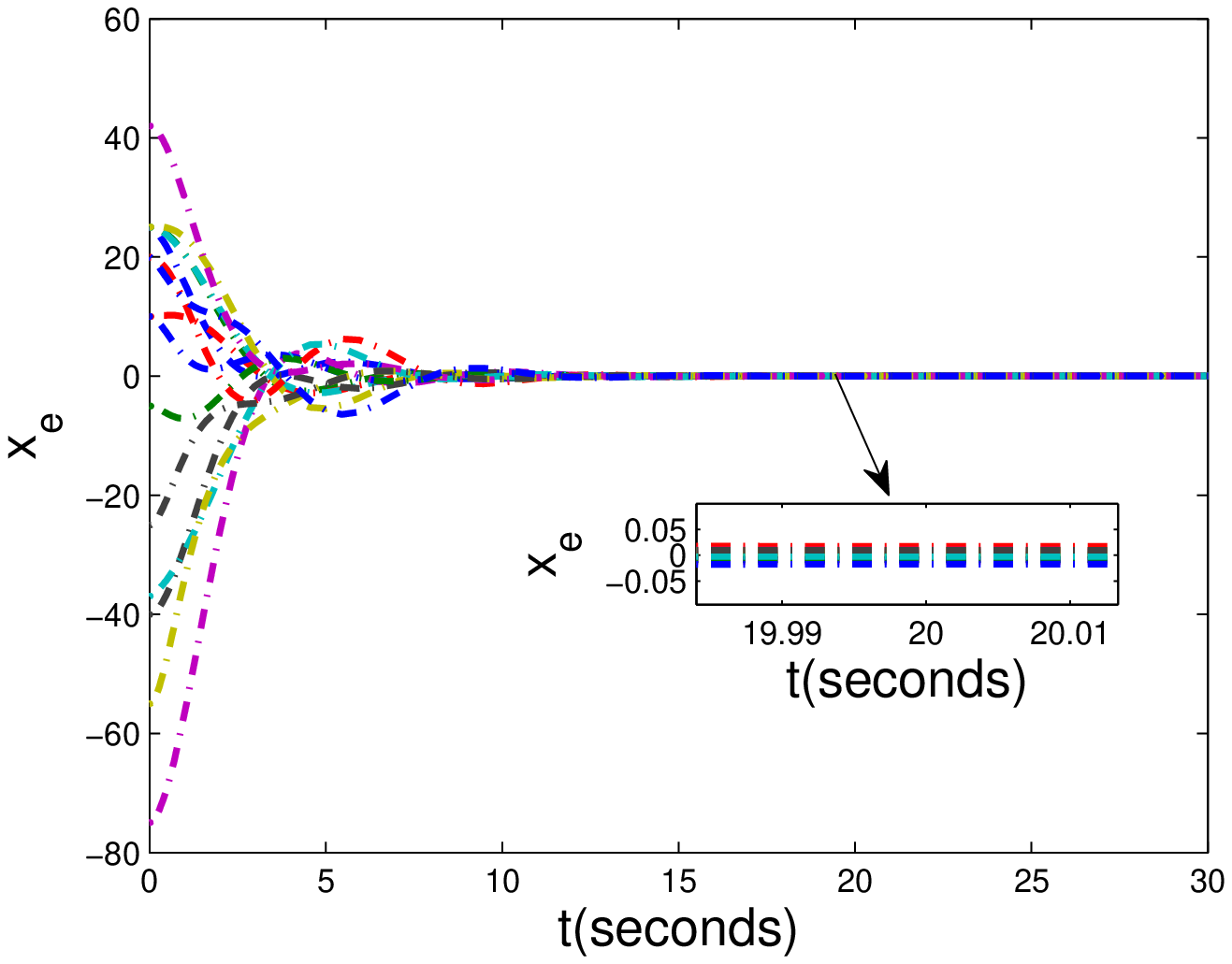}\label{fig:qlxe}}
{\includegraphics[width=72mm]{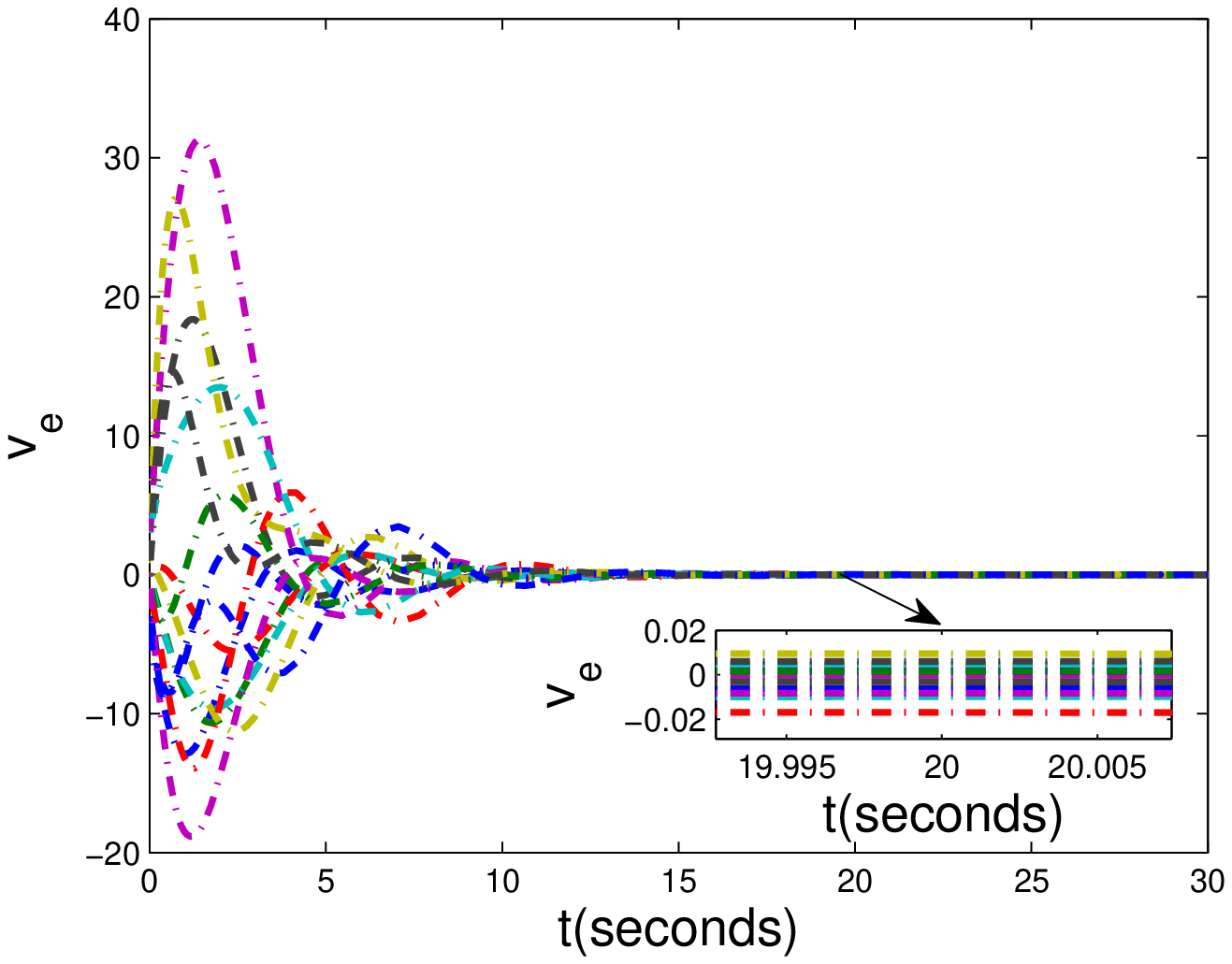}\label{fig:qlve}}
}
\caption{Edge agreement under logarithmic quantizer with $\delta_l = 0.01$.}
\label{fig:ql}
\end{center}
\end{figure*}


\begin{figure}[hbtp]
\centering
{\includegraphics[width=85mm]{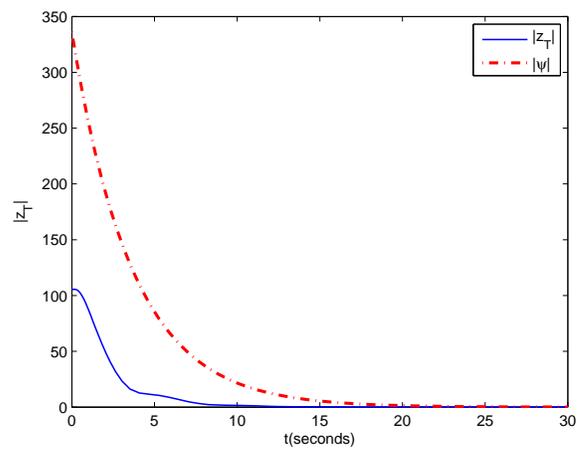}}
\caption{The convergence estimation of $\left|{{z_{_{\mathcal T}}}(t)}\right|$.}
\label{qlest}
\end{figure}

\section{Conclusions}\label{sec:conclusion}
In this paper, we presented an important concept about the essential edge Laplacian and also obtained a reduced model of the second-order edge agreement dynamics across the spanning tree subgraph. Under the edge agreement framework,  the synchronization of second-order nonlinear multi-agent systems under quantized measurements was studied. We revealed the explicit mathematical connection of the quantized interval and the convergence properties for both uniform and logarithmic quantizers. Specifically, we obtained the upper bound of the radius of the agreement neighbourhood for uniform quantizers, which indicates that the radius increases with the quantization interval. While for logarithmic quantizers, we pointed out that the agents converge exponentially to the desired agreement equilibrium. In addition, we also provided the estimates of the convergence rate.
\bibliographystyle{elsarticle-num}
\bibliography{IETCTA}

\end{document}